\documentclass[twocolumn,twoside,slac_two]{revtex4}
\usepackage{graphicx}
\usepackage{fancyhdr}
\begin{document}

\title{Observations of Gamma Ray Bursts with AGILE}

%

%
\author{Longo, F., Barbiellini, G.}
\affiliation{Dip. di Fisica, Universita' di Trieste, Trieste, Italy\\
INFN, Trieste (Italy)}
\author{Del Monte, E.}
\affiliation{IASF-INAF, Roma (Italy)}
\author{Marisaldi, M., Fuschino, F.}
\affiliation{IASF-INAF, Bologna (Italy)}
\author{Giuliani, A.}
\affiliation{IASF-INAF, Milano (Italy)}
\author{on behalf of the AGILE team}

\begin{abstract}

The AGILE satellite, in orbit since 2007, localized up to October 2009 about 1 Gamma Ray Burst (GRB) 
per month with the hard X-ray imager SuperAGILE (18 - 60 keV) (with a rate reduced by a factor 2-3 in spinning mode) 
and is detecting around 1 GRB per week with the non-imaging Mini-Calorimeter (MCAL, 0.35 - 100 MeV). 
Up to October 2011 the AGILE Gamma Ray Imaging Detector firmly detected four GRBs in the energy band between 20 MeV 
and few GeV. In this paper we review the status of the GRBs observation
with AGILE and discuss the upper limits in the gamma-ray band of the non-detected events.

\end{abstract}

\maketitle

\thispagestyle{fancy}


\section{The AGILE satellite}

Launched on April 2007, AGILE~\cite{ref1},~\cite{ref2} is a small scientific satellite of the Italian Space Agency (ASI) operating at a low inclination (2.5$^\circ$) Low-Earth Orbit at 540 km altitude and is devoted to the observation of the sky in the hard X-ray and gamma-ray energy bands . 

The AGILE payload is composed of two co-aligned imagers: the hard X-ray monitor SuperAGILE~\cite{ref3}, sensitive in the 18-60 keV energy band with a
field of view of $\sim$ 1 sr and an angular resolution of 6 arcmin, and the Gamma Ray Imaging Detector (GRID,~\cite{ref4}), a pair-tracking telescope based on a tungsten-silicon tracker~\cite{ref5}, sensitive in the energy band from 30 MeV to few GeV with a field of view of$\sim 2.5$ sr and a point spread function ranging between 3.5$^\circ$ (at 100 MeV) and 1.5$^\circ$ (at 1 GeV). The Minicalorimeter (MCAL,~\cite{ref6}) of the gamma-ray imager, based on CsI(Tl) scintillating bars, can independently detect transient events at MeV energies, using a dedicated trigger logic acting
on several time scales spanning four orders of magnitude between 290 $\mu$s and 8 seconds~\cite{ref7}, in an almost all-Sky field of view with maximum sensitivity at an angle of roughly 90$^\circ$ with respect to the satellite boresight. 

\section{Observation of GRB with AGILE}

Both SuperAGILE and MCAL are equipped with on-board triggering algorithms (see \cite{ref8} and \cite{ref7} respectively) developed to detect short gamma-ray transient events such Gamma Ray Bursts (GRBs) and Terrestrial Gamma-ray Flashes (TGFs). Dedicated telemetry packets are introduced in the data stream to downlink the trigger information. Since the beginning of the AGILE operations, SuperAGILE and MCAL are active members of the InterPlanetary Network (IPN). On November 2009 AGILE suffered a malfunction to the reaction wheel and, after that time, the satellite is working
in a spinning operative mode, with an angular velocity of $\sim$ 0.8$^\circ$ per second around the axis pointed toward the Sun
and scanning $\sim 70$ \% of the Sky every orbit. In pointing operative mode the localization rate of SuperAGILE was $\sim$ 1
GRB per month and it is now reduced of a factor of 2 - 3 in spinning mode. The MCAL detection rate is $\sim$ 1 GRB
per week and is not affected by the spinning operative mode. 

\subsection{SuperAGILE and GRBs} 

GRB 070724B\cite{ettore}, detected when AGILE was still in its Commissioning Phase, proved the capability of SuperAGILE to localize GRB with good accuracy. 
Until the end of october 2011, SuperAGILE localized 26 (20 in pointing mode, 6 in spinning) GRBs. Moreover, SuperAGILE detects about 1-2 GRBs/month outside
the field of view and these events are provided to the 3rd IPN to find the position using the triangulation technique.

\subsection{MCAL and GRBs}

MCAL has an average detection rate of about 1 GRB/week (independent of the operation mode).
Most of these detections have been independently confirmed by other instruments. As SuperAGILE, also MCAL
contributes to the IPN. The onboard trigger logic lets the satellite to download data in photon-by-photon mode after a valid trigger is issued, allowing subsequent
on-ground spectral and timing analysis of MCAL data. MCAL in-flight performance and first GRB detections were previously reported\cite{martino}. 

\subsection{GRID and GRBs}

Until October 2011 the AGILE GRID firmly detected four GRBs in the gamma-ray band: GRB 080514B~\cite{ref9},~\cite{ref10}, GRB 090401B~\cite{ref11}, GRB 090510~\cite{ref12},~\cite{ref13},~\cite{ref14} and GRB 100724B~\cite{ref15}. One more GRBs (GRB 080721) showed lower significant detection in gamma-rays. 
The four events with clear detection, show interesting features as observed by AGILE. GRB 080514B was the first gamma ray burst, since the time of EGRET, for which individual photons of energy above several tens of MeV have been detected. The hard X-ray emission observed by SuperAGILE lasted about 7 s, while there is evidence that the emission above 30 MeV extends for a longer duration (at least ~13 s). The spectral analysis of MCAL data is found to be in agreement with the measured GRID fluence. GRB090510 showed two clearly distinct emission phases: a prompt phase lasting $\sim$ 200 msec and a second phase lasting tens of seconds. The prompt phase is relatively intense in the 0.3-10 MeV range with a spectrum characterized by a large peak/cutoff energy near 3 MeV, in this phase, no significant high-energy gamma-ray emission is detected. At the end of the prompt phase, intense gamma-ray emission above 30 MeV is detected showing a power-law time decay of the flux of the type t$^{-1.3}$ and a broad-band spectrum remarkably different from that of the prompt phase. It extends from sub-MeV to hundreds of MeV energies with a photon index $\sim 1.5$. GRB 090401B shows quite an hard spectrum in the brightest peak with photon index beta $\sim 2.2 $ consistent between GRID and MCAL. Also this GRB has quite an extended emission with respect to the low energy emission. GRB100724B is the brightest burst detected in gamma-rays so far by AGILE. Characteristic features of GRB 100724B are the simultaneous emissions at MeV and GeV, without delayed onset nor time lag as shown by the analysis of the cross correlation function, and the significant spectral evolution in hard X-rays over the event duration.

The large field of view of the GRID allows to simultaneously observe about one fifth of the Sky. Motivated by the fact that significant gamma-ray emission is observed from only a small fraction of GRBs, corresponding to few events per year taking into account the AGILE and Fermi/LAT detections, we estimated the upper limits on the flux of GRBs within the GRID field of view between July 2007 and October 2009, localised by SuperAGILE, Swift/BAT, INTEGRAL/IBIS, Fermi/GBM and IPN. Our sample is composed of 68 bursts, of which 40 have spectral information,
publicly distributed through the GCN Circulars by Konus-Wind, Suzaku/WAM and Fermi/GBM.
We estimated the upper limits using a Bayesian approach and following the method proposed by \cite{ref16} and \cite{ref17}. When available, the spectral model from the publicly available information is used to calculate the flux upper limit and to extrapolate the available flux to the energy band between 30 MeV and 3 GeV. In the other cases, we adopted for these calculations the average values measured by BATSE for the photon index of an exponential cutoff and for the high energy photon index of a Band function.  We found that the calculated upper limits are constraining the extrapolation of the Band spectrum for $\sim 10$\% of the GRBs\cite{ref18}. We are still investigating the consequences of such constraints on the emission models. 
A remarkable case of this behaviour is the long GRB 090618, localised by Swift~\cite{ref19} and also by SuperAGILE (in a consistent position) and detected by the MCAL~\cite{ref20}. GRB 090618 is characterised by a remarkable fluence of ($3.2 \pm  0.6)\time 10^{-5}$ erg cm$^{-2}$, measured by MCAL in the energy band between 350 keV and 100 MeV. The GRB has a steep spectrum in the same
energy band, with a photon index of -3.16~\cite{ref20} by MCAL, and is not detected in gamma-rays by the GRID.
The interested reader can find the details and the complete analysis of the upper limits of GRBs in the GRID in a
forthcoming paper~\cite{ref18}. 

\section{High energy emission in GRBs}

The ongoing observation of GRBs with AGILE and Fermi is showing that only a small subsample of events emits in
gamma rays. In fact, the overall detection rate of both satellites is $\sim$ 10 bursts per year, consistent with current estimates~\cite{ref21}. From the data analysis it is emerging that GRBs emitting in the GeV band are also bright at lower energy. In
fact, an analysis of eleven GRBs detected by Fermi/LAT until October 2009 shows that the fluence in Fermi/GBM (8 keV - 10 MeV) of these events belongs to the highest tail of the distribution~\cite{ref22}. 
From the AGILE and Fermi observations we can see that the GeV emission of GRBs is simultaneous to the prompt
phase, often with a delayed onset and extended emission. Some events have the same spectral shape from keV up to
GeV energies, modelled by a single Band function (for example GRB 080514B, \cite{ref9}), while in other the gamma-ray
emission is fitted by additional components, as for example in GRB 090510~\cite{ref12},\cite{ref13}) or GRB 090902B below
50 keV and above 100 MeV~\cite{ref23}.
Up to now the afterglow of two GRBs detected in gamma-rays by AGILE, GRB 090401B~\cite{ref11} and GRB 090510~\cite{ref14},
has been observed by Swift.  In our analysis we investigated the GRID data of the sample of GRBs within our field of view and we did not find any significant gamma-ray afterglow emission until 3600 s after trigger~\cite{ref18}.

\section{Conclusions}

Thanks to the high level capabilities of the its different detector components, the AGILE satellite obtained several important results on GRB science.
The AGILE satellite is capable to detect GRBs in an energy range extended over 6 orders of magnitude.
AGILE detected few GRBs but representing all the main features present in GRB high energy domain, such us delayed onset, 
spectral extra-component and extended emission. The Upper limits that AGILE derived are useful to exclude 
the presence of an extra component at the same fluence level as the low energy emission in most of the GRB in its field of view.

{\it Acknowledgements.} The AGILE Mission is funded by the Italian
Space Agency (ASI) with scientific and programmatic participation by 
the Italian Institute of Astrophysics (INAF) and the Italian Institute
of Nuclear Physics (INFN).

\clearpage

\end{document}